\documentstyle[prl,aps,psfig,floats]{revtex}
\tighten

\begin{document}
\draft
\preprint{} 
\twocolumn[\hsize\textwidth\columnwidth\hsize\csname@twocolumnfalse\endcsname 
\title{A Unified treatment of small and large- scale dynamos in helical turbulence }
\author{Kandaswamy Subramanian\\
National Centre for Radio Astrophysics, TIFR, 
Poona University Campus, \\ 
Ganeshkhind, Pune 411 007, India. }
\maketitle

\begin{abstract}

Helical turbulence is thought to provide the key to
the generation of large-scale magnetic fields.
Turbulence also generically leads to rapidly growing small-scale 
magnetic fields correlated on the turbulence scales.
These two processes are usually studied separately.
We give here a unified treatment of both
processes, in the case of random fields, 
incorporating also a simple model non-linear drift.
In the process we uncover an interesting plausible
saturated state of the small-scale dynamo and  
a novel analogy between quantum mechanical (QM) 
tunneling and the generation of large scale fields. 
The steady state problem of 
the combined small/large scale dynamo, is 
mapped to a zero-energy, QM potential 
problem; but a potential which, for
non-zero mean helicity, allows tunneling of bound states.  
A field generated by the small-scale 
dynamo, can 'tunnel' to produce large-scale correlations,
which in steady state, correspond to a force-free 'mean' field.

\end{abstract}

\maketitle
\date{\today}
] \renewcommand{\thefootnote}{\arabic{footnote}} \setcounter{footnote}{0}

Large-scale magnetic fields in astronomical objects
are thought to be generated by
dynamo action involving helical turbulence 
and rotational shear \cite{dynam,moff}.
Here large-scale refers to scales much larger than the 
outer scale, say L, of the turbulence. 
However, turbulent motions, with a
large enough magnetic Reynolds number (MRN henceforth),
can also excite a small-scale dynamo, which exponentiates 
fields correlated on the tubulent eddy scale, 
at a rate much faster than the mean field
growth rate \cite{dynam,ssd}. These two dynamo problems, 
viz. the small-scale dynamo (SSD) and large-scale dynamo (LSD), 
are usually treated separately. 
However this separation is often artificial; there is
no abrupt transition from the field correlated
on scales smaller than $L$ and that correlated on
larger scales. We show here that the equations for the
magnetic correlations, which involve both the
longitudinal and helical parts, are already sufficiently general to
incorporate both small and large-scale dynamos in the
case of random fields. They provide us with a paradigm to study 
the dynamics in a unified fashion, which could be
particularly useful to study possible inverse
cascade of magnetic fields, to scales larger than $L$.

Consider the induction equation for the magnetic field.
\begin{equation}
(\partial {\bf B}/ \partial t) =
{\bf \nabla } \times ( {\bf v} \times {\bf B} - 
 \eta {\bf \nabla } \times {\bf B}), 
\label{basic}
\end{equation} 
where ${\bf B}$ is the magnetic field,  
${\bf v}$ the velocity of the fluid and
$\eta$ the ohmic resistivity. 
Take ${\bf v} = {\bf v}_T + {\bf v}_D$, the
sum of an externally prescribed stochastic
field ${\bf v}_T$, and a drift component ${\bf v}_D$, 
which models the non-linear back reaction of the
growing Lorentz force.
We assume ${\bf v}_T$ to be an isotropic, homogeneous,
Gaussian random velocity field with 
zero mean. For simplicity, we also assume ${\bf v}_T$ to have 
a delta function correlation in time (Markovian approximation). 
Its two point correlation is specified as 
$<v^i_T({\bf x},t)v^j_T({\bf y},s) > = T^{ij}(r) \delta (t-s)$ 
with (cf. \cite{landau})
\begin{equation}
T^{ij}(r) =
T_{N}[\delta^{ij} -({r^i r^j \over r^2})] +
 T_{L}({r^i r^j \over r^2}) +
C\epsilon_{ijf} r^f .
\end{equation}
Here $<>$ denotes ensemble averaging, 
$r = \vert {\bf x} -{\bf y} \vert$ and $r^i = x^i -y^i$.
$T_{L}(r)$ and $T_{N}(r)$ are the 
longitudinal and transverse correlation functions for the velocity 
field while $C(r)$ represents 
the helical part of the velocity correllations.
If ${\bf \nabla}.{\bf v}_T = 0$,
$ T_{N} = (1/ 2r) d(r^2 T_{L})/dr$. 
(Note ${\bf v}_T$ is also not correlated with the magnetic field.)

To model the drift velocity in a tractable manner, but one which
nevertheless gives some feel for possible nonlinear effects,
we proceed as follows. As the
magnetic field grows, the Lorentz force pushes on
the fluid. We assume the fluid almost
instantaneously responds to this push, and 
develops an extra, 'drift' component to the velocity
proportional to the instantaneous Lorentz force. So we take a model 
${\bf v}_D = a[({\bf \nabla } \times {\bf B}) \times {\bf B})]$,
with the parameter $a = \tau / (4\pi \rho)$, where
$\tau$ is some response time, and $\rho$ is the fluid density.
(Such a velocity can also arise when friction dominates
inertial forces on ions, as in ambipolar drift).
This gives a model nonlinear problem, where the nonlinear
effects of the Lorentz force are taken into account as
simple modification of the velocity field. 
Such a phenemenological modification of the velocity field has in fact been
used by Pouquet {\it et al.} \cite{pouq} and 
Zeldovich {\it et al.} (pg. 183) \cite{dynam} to discuss 
nonlinear modifications to the alpha effect, and we adopt it below.

Consider a system, whose size $S >> L$,
and for which the mean field averaged
over any scale is zero. 
Of course, the concept of a large-scale field still
makes sense, as the correlations between field components
separated at scales $r >> L$, can in principle be non-zero.
We take ${\bf B}$ to be a homogeneous,
isotropic, Gaussian random field with zero mean. This
is a perfectly valid assumption to make in the kinematic
regime (${\bf v}_D = 0$), as the stochastic ${\bf v}_T$, 
has these symmetries. When we include the non-linear drift
velocity, it amounts to making a closure hypothesis,
which we do here again for analytical tractability.
The equal-time, two point correlation of the magnetic field is given by 
$<B^i({\bf x},t) B^j({\bf y},t) > = M^{ij}(r,t)$, where 
\begin{equation}
M^{ij} = M_N[\delta^{ij} -({r^i r^j \over r^2})] + 
M_L ({r^i r^j  \over r^2}) + H \epsilon_{ijf} r^f .
\label{mcor}
\end{equation}
(Here $<>$ denotes a double ensemble average, over both the 
stochastic velocity and stochastic ${\bf B}$ fields).
$M_L(r,t)$ and $M_N(r,t)$ are the longitudinal 
and transverse correlation functions for the magnetic field 
while $H(r,t)$ represents 
the (current) helical part of the correlations.
Since ${\bf \nabla}.{\bf B}=0$, 
$M_N = (1/ 2r) \partial (r^2 M_{L})/ (\partial r)$. 

The stochastic Eq.\ (\ref{basic}) can be 
converted into the evolution equations for $M_L$ and $H$. 
We give a detailed derivation of these equations 
elsewhere, including the effect of the non-linear drift \cite{subamb}.
We get 
\begin{equation}
{\partial M_L \over \partial t} = {2\over r^4}{\partial \over \partial r}
(r^4 \kappa_N {\partial M_L \over \partial r}) + G M_L - 4\alpha_N
H  \label{mleq}
\end{equation}
\begin{equation} {\partial H\over \partial t} = 
{1\over r^4}{\partial \over \partial r}
\left(r^4  {\partial \over \partial r}(2\kappa_N H 
+ \alpha_N M_L)\right) 
\label{mheq}
\end{equation}
where we have defined 
\begin{eqnarray}
&&\kappa_N = \eta + T_{L}(0) - T_{L}(r) + 2aM_L(0,t) \nonumber\\ 
&&\alpha_N = 2C(0) - 2C(r) -4aH(0,t) \nonumber\\
&&G = - 4\left[(T_{N}/ r)^{\prime} + (rT_{L})^{\prime}/r^2
\right]
\label{coeff}
\end{eqnarray}
Here prime denotes a derivative with respect to $r$.
These equations form a closed 
set of nonlinear partial differential
equations for the evolution of $M_L$ and $H$,
describing, as we will see, both SSD and LSD action
for random fields. The effective diffusion $\kappa_N$ includes 
microscopic diffusion ($\eta$), a scale-dependent 
turbulent diffusion ($T_{L}(0)-T_{L}(r)$) and 
nonlinear drift adds an amount $ 2aM_L(0,t)$, proportional 
to the energy density in the fluctuating fields.
Similarly $\alpha_N$ 
represents a scale dependent $\alpha$-effect ($2C(0) -2C(r)$) and 
nonlinear drift decreases this by $4aH(0,t)$, 
proportional to the mean 
current helicity of the magnetic fluctuations.
This modification to the $\alpha$-effect is
the same as that obtained in \cite{dynam,pouq}.
The $G(r)$ term allows for 
the rapid generation of magnetic fluctuations by velocity shear 
and the existence of a SSD {\it independent} of any 
large-scale field \cite{dynam,ssd}.

First consider non-helical turbulence, with $C(r) = 0$, 
allowing solutions with $ H(r,t) = 0$. This 
case has been extensively studied 
for the kinematic case ($a=0$), when $T_L(r)$ has
a single scale (cf. \cite{dynam,ssd}) and by us 
for a model Kolmogorov type turbulence \cite{subamb}. 
These studies show
that one has SSD action, and magnetic fields correlated
on scales upto the turbulent scale can be generated.
In the kinematic limit, $\kappa_N$ and $G$ 
are time independent. One can then look
for eigenmode solutions to (\ref{mleq}),
of the form 
$\Psi(r)\exp(2\Gamma t)  =r^2\sqrt{\kappa_N}M_L$.
This transforms Eq. (\ref{mleq}) for $M_L(r,t)$, 
into a time independent,
Schrodinger-type equation,
but with a variable (and positive) mass, 
\begin{equation}
-\Gamma \Psi  = 
-\kappa_N{d^2 \Psi \over d r^2} + U_0(r)\Psi . 
\label{sievol}
\end{equation}
The "potential" is
$U_0(r) =  T_{L}^{\prime\prime} + (2T_{L}^{\prime}/r) 
+ \kappa_N^{\prime\prime}/2
-(\kappa_N^{\prime})^2/(4\kappa_N) + 2\kappa_N/r^2$, for a
divergence free velocity field.
The boundary condition is $\Psi \to 0$, as $r \to 0, \infty$.
Note that $U_0 \to 2\eta/r^2$ as $r \to 0$,
while $U_0 \to 2(\eta + T_L(0))/r^2$ as $ r\to \infty$.
The possibility of growing modes with $\Gamma > 0$ 
obtains, if one can have a potential well, with 
$U_0 $ sufficiently negative in some range of $r$, 
to allow the existence of {\it bound states}, 
with an "energy" $E = -\Gamma < 0$.

Suppose we have turbulent motions on a single scale $L$, 
with a velocity scale $v$. Define the magnetic Reynolds 
number (MRN) $R_m = vL/\eta$. Then one finds \cite{ssd,subamb} 
that there is a critical MRN, $R_m = R_c \approx 60$, 
so that for $R_m > R_c$, the potential $U_0$ allows the 
existence of bound states. For $R_m = R_c$, $\Gamma = 0$, 
and this marginal stationary state is the "zero" energy 
eigenstate in the potential $U_0$ \cite{poten}.  For $R_m > R_c$, 
$\Gamma > 0$ modes of the SSD can be excited, and 
the fluctuating field correlated on a 
scale $L$, grows exponentially, on the
corresponding 'eddy' turn-over time scale, with a
growth rate $\Gamma_L \sim v/L$ \cite{kol}.
To understand the spatial structure of the fields, 
define $w(r,t) = < {\bf B}({\bf x},t) . {\bf B}({\bf y},t)>  
 = (1/ r^2) d[r^3 M_L]/dr$, 
the correlated dot product of the random field. 
For the SSD, $w(r)$ is strongly peaked 
within a region $r =r_d \approx L (R_m)^{-1/2}$ about the origin, 
for all the modes, and for the fastest growing mode, 
changes sign accross $r \sim L$ and rapidly decays
with increasing $r/L$. 
(For the marginal mode with $\Gamma =0$, $R_m$ is
replaced by $R_c$). Note that $r_d$ is the diffusive scale 
satisfying the condition $\eta/r_d^2 \sim v/L$. 
A pictorial intepretation of the correlation function, due to the 
Zeldovich school (cf. \cite{dynam,ssd}), is to think of the field 
as being concentrated in "flux ropes" with 
thickness of order $r_d << L$, 
and curved on a scale up to $\sim L$,
to account for negative values of $w$.

How does the SSD saturate? 
The back reaction, in the form of a
non-linear drift, simply replaces $\eta$ by an
effective, time dependent $\eta_D = \eta + 2aM_L(0,t)$, in
the $\kappa_N$ term of Eq. (\ref{mleq}).
Suppose we define an effective MRN, for
fluid motion on scale $L$, by $R_D(t) = v L/ \eta_D(t)$.
Then as the energy density in the fluctuating field,
say $E_B(t) = 3M_L(0,t)/8\pi$,
increases, $ R_D$ decreases.
In the final saturated state, 
with $(\partial M_L/\partial t) =0$ 
(obtaining say at time $t_s$),
$M_L$, and hence the effective $\eta_D$ in (\ref{mleq}) 
become independent of time.
Solving for this stationary state then becomes 
identical to solving for the marginal (stationary) mode of the
kinematic problem, except that $R_m$ is replaced by $R_D(t_s)$. 
The final saturated state is then the marginal eigenmode which obtains,
when $E_B$ has grown (and $R_D$ decreased) such that
$R_D(t_s) = v L/( \eta + 2aM_L(0,t_s) ) = R_c \sim 60$. 
Also $w(r)$ for the saturated state will be strongly peaked 
within a region $r =r_d \approx L (R_c)^{-1/2}$ about the origin, 
change sign accross $r \sim L$ and then rapidly decay for larger $r/L$.
From the above constraint, 
$M_L(0,t_s) = vL /(2a R_c)$, assuming 
$\eta << 2aM_L(0,t_s)$. So at saturation, 
\begin{equation}
E_{B}(t_s) = {3M_L(0,t_s) \over 8\pi} = {3 \over 2} 
{\rho v^2 \over 2} {L/v \over \tau} {1 \over R_c}
\label{saten}
\end{equation}
Note that $\tau$ is an unknown model parameter. If we were 
to adopt $\tau \sim L/v$, that is the eddy turn-over time,
then $E_B$ at saturation is a small fraction $\sim R_C^{-1} \ll 1$,
of the equipartition value.
Further suppose we intepret $w(r)$ for the saturated state in terms
of the Zeldovich {\it et al.} picture; of flux ropes of thickness $r_d$,
curved on scale $L$, in which a field of strength $B_p$ is concentrated.
In this picture, the average energy density in the field
$E_B \sim (B_p^2/8\pi) L r_d^2/L^3 $. Using $r_d^2/L^2 \approx R_c^{-1}$,
and $\tau \sim L/v$, we then have $B_p^2/8\pi \sim \rho v^2/2 $,
where, remarkably, the $R_C^{-1}$ dependence has disappeared, 
and $B_P$ has equipartition value.
So the SSD could saturate with
the small-scale field
having peak values of order the equipartition field, 
being concentrated into flux ropes of 
thickness $LR_c^{-1/2}$, curved on scale $L$, 
and an average energy density $R_c^{-1}$
times smaller than equipartition.

We now turn to consider the effect of
helical correlations. If $\alpha_0 = 2C(0) \ne 0$, then 
one can see from Eq. (\ref{mleq}) and (\ref{mheq}), that 
new generation terms arise at $r >> L$, due to
the $\alpha$- effect, in the form 
$\dot M_L = .... -4\alpha_T H$ and $\dot H = ... + 
 [r^4 (\alpha_T M_L)^{\prime}]^{\prime}/r^4 $.
Here $\alpha_T = \alpha_0 - 4aH(0,t)$ and 
dot represents a time derivative.
These couple $M_L$ and $H$
and lead to the growth of large-scale correlations.
There is also decay of the correlations at $r \gg L$, due to 
diffusion with an effective diffusion
co-efficient, $\eta_T = \eta + T_{L}(0) + 2a M_L(0,t)$.
From dimensional analysis, 
the effective growth rate is $ \Gamma_D \sim 
\alpha_T/D - \eta_T/D^2$, for
correlations on scale $\sim D$, as in the large-scale
$\alpha^2$- dynamo. This also picks out a special scale 
$D_0 \sim \eta_T/\alpha_T$ for a stationary state
(see below). Further, as the SSD, is simulataneously
leading to a growth of $M_L$ at $r < L$, 
in general at a faster rate $v/L >> \alpha_0/D$,
the growth of large-scale correlations can 
be seeded by the tail of the SSD eigenfunction at $r > L$.
The SSD generated small-scale field can thus seed the
large-scale dynamo. Indeed, as advertised, both
the SSD and LSD operate simulataneously when $\alpha_0 \ne 0$, 
and can be studied simply by solving for one $M_L(r,t)$.

The coupled time evolution of $H$ and $M_L$ for a non-zero 
$\alpha_0$  requires numerical solution. But interesting analytical 
insight into the system can be obtained for the marginal,
stationary mode, with
$(\partial M_L/\partial t) =(\partial H/\partial t) =0$. 
Infact both the kinematic and non-linear dynamo problem can be treated in
a unified fashion.
With $H$ independent of time, Eq. (\ref{mheq}) implies 
$2 \kappa_N H + \alpha_N M_L = 0$, for any solution regular
at $r=0$ and vainishing at $r \to \infty$. From this, 
as $r\to 0$, $2 \eta H(0,t) = 0$, 
and hence $H(0,t) = 0$ for a non-zero $\eta$,
a result which also follows directly from 
the evolution (conservation) of magnetic helicity \cite{hel}.
So any general nonlinear addition to the $\alpha$-effect which arises
in terms of $H(0,t)$ has to vanish in a stationary state!
\begin{figure}
\begin{picture}(200,180)
\psfig{figure=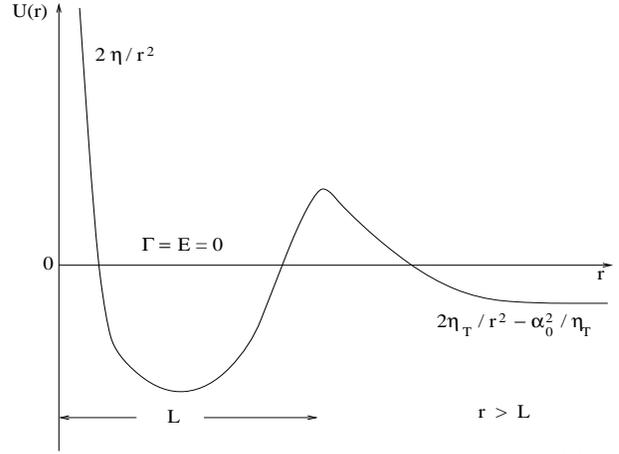,height=6.0cm,width=8.0cm,angle=270.0}
\end{picture}
\caption{ Schematic illustration of the potential
$U(r)$ for the marginal mode in helical turbulence.
A non zero $\alpha_0$ allows the tunneling
of the zero-energy state to produce
large-scale correlations.}
\end{figure}

Now substitute $H(r) = - \alpha_N(r) M_L(r)/ (2 \kappa_N(r))$
into (\ref{mleq}) and define once again 
$\Psi =r^2\sqrt{\kappa_N}M_L$. We get 
\begin{equation}
-\kappa_N{d^2 \Psi \over dr^2} + \Psi \left [
U_0 - {4 [ C(0) - C(r) ]^2 \over \kappa_N} \right] = 0.
\label{sievoln}
\end{equation}
We see that the problem of determining the magnetic
field correlations, for the marginal/stationary mode
once again becomes the problem of determining
the zero-energy eigen-state
in a modified potential, $U = U_0 - 4 [C(0) - C(r)]^2/\kappa_N$.
Note that the addition to $U_0$, due to the helical correlations,
is always negative definite. So helical correlations
tend to make bound states easier to obtain.
When $C(0) = 0$, and there is no net $\alpha$-effect,
the addition to $U_0$ vanishes at
$r \gg L$, and $U \to 2\eta_T/r^2$ at large $r$,
as before. The critical MRN for the stationary state,
will however be smaller than when $C(r) \equiv 0$,
because of the negative definite addition to $U_0$.

When $\alpha_0 = 2C(0) \ne 0$, a remarkable change occurs
in the potential. At $r \gg L$, where the turbulence
velocity correlations vanish, we have
$U(r) = 2\eta_T/r^2 - \alpha_0^2/\eta_T$.
So the potential $U$ tends to a negative definite constant
value of $ - \alpha_0^2/\eta_T$ at large $r$ (and
the effective mass, $1/2\kappa_N \to 
1/2\eta_T$, a constant with $r$.) 
There are strictly no bound states, 
with zero energy/growth rate, 
for which the correlations
vanish at infinity. 
We have schematically illustrated the resulting 
potential $U$ in Figure 1, which is a 
modification of figure 8.4 of 
Zeldovich {\it et al.} \cite{dynam}. In fact, for a 
non-zero $\alpha_0$, $U$ corresponds
to a potential which allows tunneling (of the bound state) 
in the corresponding quantum mechanical (QM) problem. 
It implies that the correlations are necessarily
non-zero at large $r > L$. The analytical solution
to (\ref{sievoln}) at large $r \gg L$, 
is easily obtained. We have for $r \gg L$, $ M_L(r) = {\bar M}_L(r)$, 
\begin{equation}
{\bar M}_L(r) = 
{1 \over r^{3/2}} [ C_1 J_{3/2}(\mu r) + C_2 J_{-3/2}(\mu r)],
\label{larb}
\end{equation}
where $ \mu = \alpha_0/\eta_T = D_0^{-1}$, and $C_1,C_2$ arbitrary
constants.  Also 
$w(r) = {\bar w}(r)= \mu r^{-1}[C_1 \sin \mu r  + C_2 \cos \mu r ]$.
Clearly for a non-zero $\alpha_0$, the correlations 
in steady state at large $r$, are like "free-particle" states, 
extending to infinity! 

An alternate derivation of $\bar M_L(r)$, 
for the kinematic case, clarifies its meaning further.
Suppose one thinks of the 
large-scale field as a "mean" field 
${\bf B}_0$, the mean taken over cells much larger than
$ L$. Assume ${\bf B}_0$ itself is random over
different cells, statistically homogeneous and isotropic,
with a correlation $< B_{0i} B_{0j} > = M_{ij}^L(r)$.
Let $M_L^L$, $M_N^L$ and $H^L$ be respectively the
corresponding longitudinal, transverse and helical correlations.
Then ${\bf B}_0$ in each cell
obeys the kinematic, mean-field dynamo equation 
$(\partial {\bf B}_0/ \partial t) =
{\bf \nabla } \times ( \alpha_0 {\bf B}_0 - 
(\eta + T_{L}(0)) {\bf \nabla } \times {\bf B}_0)$,
whose steady state solution is, 
${\bf \nabla } \times {\bf B}_0 = \mu_0 {\bf B_0}$,
where $\mu_0 = \mu (a=0)$.
This constraint, imposed on $M_{ij}^L$, 
gives $H^L = -\mu_0 M_L^L/2$, and
$(M_L^L - M_N^L)/r^2 - M_N^{\prime}/r = \mu_0^2 M_L/2$.
Also as ${\bf \nabla}.{\bf B}_0=0$, 
$M_N^L = (1/ 2r) (r^2 M_{L})^{\prime}$. 
These three equations fix all the functions uniquely.
We get, remarkably, $M_L^L(r) = {\bar M}_L(r)$,
with $\mu$ replaced by the kinematic value $\mu_0$.
So this solution actually describes a random mean-field, 
for the marginal large-scale
dynamo. Similarly if we had imposed  
${\bf \nabla } \times {\bf B}_0 = \mu {\bf B_0}$, 
$M_L^L$ would be given by Eq.(\ref{larb}). Note this also shows that
the effective, steady-state, large-scale field ${\bf B_0}$ is force-free, 
although ${\bf B}$ itself is not.

It is straight-forward to connect the large-scale,
force-free field for the
marginal mode of helical turbulence, 
with the SSD generated field, as they are both
the solution of the same Eq. (\ref{sievoln}),
for large and small $r$ respectively. 
For example, 
one can integrate Eq. (\ref{sievoln}),
adopting different starting values of $M_L(0,t_s)$,
and taking $M_L^{\prime}(0,t_s) = 0$, to construct a whole family
of solutions (parameterised by $M_L(0,t_s)$), 
which match small-scale correlations with
the large-scale correlation of (\ref{larb}).
For each such solution, we will have one 
value of $C_1/C_2$.
Note that this is unlike
the standard QM tunneling problem, where the boundary
condition that the free-particle state is an outgoing
wave at large $r$ uniquely fixes the tunneling amplitude,
for a given bound state. However,
when we consider a zero-energy, stationary state,
there is no such natural time-asymmetric boundary condition;
so no unique fix for $C_1/C_2$, in (\ref{larb}). 
Nevertheless, if $M_L(0,t_s)$ is so small,
and $R_D$ so large, 
that $U$ admits bound states with
energy $E < -\alpha_0^2/\eta_T$, then the corresponding
time dependent system is unlikely to lead to 
stationary correlations.
This sets a lower bound on $M_L(0,t_s)$ or $E_B$.
Further as $\dot M_L(0,t) = ..... 
 + 16H^2(0,t)$, the saturation of $M_L(0,t)$,
depends on how fast $H(0,t) \to 0$, its stationary value;
over and above the saturation
effects of the increasing diffusion $\eta_D$. So
the full time dependent problem needs to be solved
to fix an upper bound on $M_L(0,t_s)$ or $E_B$ \cite{tim}.
We will return to this elsewhere.

In conclusion, we have given here 
a unified treatment of small- and large-scale dynamos,
for the case of random fields, 
incorporating also a simple model non-linear drift.
We uncovered an interesting plausible
saturated state of the small-scale dynamo.
For random fields, we argued that any non-linear addition to the
$\alpha$ effect in terms of the average current helicity
$H(0,t)$ (cf.\cite{dynam,pouq}), has to vanish in a steady state.
The steady state problem of 
the combined small/large scale dynamo, was then 
mapped to a zero-energy, QM potential 
problem; but a potential which, for
$\alpha_0 \ne 0$, allows tunneling of bound states.  
A field generated by the SSD, 
can then "tunnel" to seed the growth of  
large-scale correlations,
which in steady state, correspond to a force-free "mean" field.
It remains to solve the fully time-dependent system
and to incorporate more realistic back-reaction
effects of the Lorentz force. 

{\it Acknowledgments:} KS thanks the University of Newcastle
for hospitality and Axel Brandenburg, T. Padmanabhan, Anvar Shukurov, 
Dmitry Sokoloff and S. Sridhar for insightful discussions.


\begin{references}
\bibitem{kol} In Kolmogorov type turbulence,
the mode correlated on the cut-off scale grows fastest.
But modes with extent upto the outer scale $L$ can also grow at
their corresponding, slower eddy-turn over rates
\cite{subamb,ssd}.
\bibitem{hel} From (\ref{mleq}) and (\ref{mheq}),  
$ (dI_M/dt) = -2\eta < {\bf B}.({\bf \nabla} \times 
{\bf B}) > = 12 \eta H(0,t)$, where $I_M = <{\bf A}.{\bf B}>$ 
is proportional to the magnetic helicity
and ${\bf A}$ is the vector potential. 
So as $\eta \to 0$, $I_M$ is
is conserved in the ideal limit, as required \cite{moff}. 
But for stationary states with $dI_M/dt = 0$, 
when $\eta \ne 0$, then $H(0,t) = 0$.
\bibitem{tim} Solving the time dependent system
will also clarify how initially localised correlations, 
``propogate'' to larger and larger scales, due to the $\alpha_T$
coupling, and asymptotically reach the large $r$ stationary state 
(viz. Eq. (\ref{larb})).
\end{references}
\end{document}